\documentclass[12pt]{article}
\usepackage{epsf,axodraw}
\newcommand{\lsim}{\stackrel{<}{_\sim}}
\newcommand{\gsim}{\stackrel{>}{_\sim}}

\newcommand{\nn}{\nonumber}

\def\gsim{\ \rlap{\raise 3pt \hbox{$>$}}{\lower 3pt \hbox{$\sim$}}\ }
\def\lsim{\ \rlap{\raise 3pt \hbox{$<$}}{\lower 3pt \hbox{$\sim$}}\ }
\begin{document}

\thispagestyle{empty}
\begin{flushright}
\rightline{UAB-FT-514}
\rightline{hep-ph/0105103}
\rightline{May 2001}
\end{flushright}

\vspace*{0.8cm}
\centerline{\Large\bf 
Model independent sum rules
}
\vspace*{0.2cm}
\boldmath 
\centerline{\Large\bf 
for $B \rightarrow \pi K$ decays
}
\unboldmath
\vspace*{0.2cm}

\vspace*{1.5cm}
\centerline{{\sc Joaquim Matias
}}

\vspace*{1cm}
\centerline{\sl Institut f\"ur Theoretische Physik E, RWTH 
Aachen,
D - 52056 Aachen, Germany
\footnote{
 supported in part by BMBF,
contract 05 HT9PAA 1                            
}}
\centerline{\sl and}
\centerline{\sl IFAE, Universitat Aut\`onoma de Barcelona, Spain} 
\vspace*{0.2cm}

\vspace*{0.9cm}
\centerline{\bf \large Abstract}
\vspace*{0.8cm}

\noindent We provide  a set of sum rules relating CP-averaged 
branching ratios and CP-asymme-tries of the
$B \rightarrow \pi K$ modes. They
prove to be useful as a mechanism to `test' experimental data given
 our expectations of the size of isospin
breaking. A set of observables emerges providing a simpler 
interpretation of data in terms of isospin breaking. Moreover, the 
derivation is done in a completely model
 independent way, i.e., they can accommodate also 
New Physics contributions.

\noindent 
\vfill
\leftline{May 2001}    
\newpage
\centerline{}
\thispagestyle{empty}
\setcounter{page}{1}    
\newpage
\pagenumbering{arabic}


$B$ physics  offers the
possibility to test the mechanism of CP violation in the Standard Model,
 i.e., a
single phase in the
quark flavor mixing matrix
but also to seek for the first clues of
New Physics. Non-leptonic $B$ meson decays, in particular, 
$B \rightarrow \pi K$ modes are
nicely
suited for both purposes. These modes will play a central role in the
 determination of the weak
angle $\gamma$ of the Unitarity Triangle (UT). There has been 
an intense activity to improve the theoretical predictions, 
following different strategies to control their 
hadronic uncertainties and to extract the relevant
 information, i.e., $\gamma$ and
the strong phases. One of the main differences between these strategies
 is the way to deal with 
strong phases, either 
using $SU(3)$ together with other experimental data  \cite{fl,fl2,n1}, 
or predict them from first principles
 using QCD factorization \cite{bbns99,bbns99bis,bbns00,mu}, or using 
Wick 
contractions to 
combine factorization with a
parametrization of the penguin amplitude
 as in \cite{silv}.
In the present work, we will follow the QCD factorization approach 
 also taking into account the impact of annihilation 
topologies \cite{bbns00}.  (The
potential importance of annihilation was also noted in \cite{keum}.)

The angle $\gamma$ of the 
UT
obtained from these modes should be compared with other determinations 
of $\gamma$, if a different
value is found this would signal New Physics \cite{jo,neu3}.
There has been a considerable effort in the last years to find  
strategies \cite{fl,fl2,gr3} to constrain the angle  $\gamma$.

On the experimental side, $B$ factories have  reported recently 
new data 
\cite{exp,exp2} on 
 the set of  charged
 and neutral non-leptonic decay modes:
$B
\rightarrow \pi K$. 
The collected data on the branching ratios of these decay modes are
organized in two types of observables. A first type of observables are 
the CP-averaged branching ratios 
\cite{fl2,n1}:
\begin{eqnarray} \label{eqa}
R_{\;}&=&{
 } \left[{{\rm BR}(B_{d}^0 
\rightarrow \pi^- K^+) + {\rm
BR}({\overline{ B_{d}^0}}
\rightarrow
\pi^+ K^- )
\over {\rm BR}(B^+ \rightarrow \pi^+ K^0) + 
{\rm BR}(B^- \rightarrow \pi^- {\overline{ K^0} }
)}
\right], \nn
 \\ \label{eqa2}
R_{c}&=&{2} \left[{
{\rm BR}(B^+ \rightarrow \pi^0 K^+) + {\rm BR}(B^- \rightarrow
\pi^0 {K^-} ) 
\over 
{\rm BR}(B^+ \rightarrow \pi^+ K^0) + {\rm
BR}(B^- \rightarrow
\pi^- {\overline{ K^0}} )      
}
\right], \nn
 \\ \label{eqa3}
R_{0}&=&{2 {
 }} \left[{{\rm BR}(B_{d}^0 
\rightarrow \pi^0 K^0) + 
{\rm
BR}(\overline{B_{d}^0} \rightarrow
\pi^0  \overline{K^0} )
\over {\rm BR}(B^+  \rightarrow \pi^+ K^0) + {\rm 
BR}(B^- \rightarrow
\pi^- \overline{K^0} )}
\right].                                                                             
\end{eqnarray}
We prefer to use these definitions because, in terms of them,
 the expressions for the sum rules become  simpler.
Other definitions for the charged and neutral channels that are used 
in the literature are
$R_{*}=1/R_{c}$ \cite{n1} and $R_{n}=R/R_{0}$ \cite{fl2}. 
 CP-asymmetries are the second 
type of observables:
\begin{eqnarray} \label{eqabis}
{\cal A}_{\rm CP}^{0 +}&=&{{\rm BR}( B^+ \rightarrow \pi^0 K^+ )-
{\rm BR}( B^- \rightarrow \pi^0 { K^-} ) \over
{\rm BR}( B^+ \rightarrow \pi^0 K^+ )+ {\rm BR}( B^- \rightarrow \pi^0 {
K^- }
)}, 
\nn \\
{\cal A}_{\rm CP}^{+ 0}&=&{{\rm BR}( B^+ \rightarrow \pi^+ K^0 )-
{\rm BR}( B^- \rightarrow \pi^- \overline{ K^0} ) \over
{\rm BR}( B^+ \rightarrow \pi^+ K^0 )+ {\rm BR}( B^- \rightarrow \pi^-
\overline{K^0 }
)},
\nn \\      
{\cal A}_{\rm CP}^{- +}&=&{{\rm BR}( B_{d}^0 \rightarrow \pi^- K^+ )-
{\rm BR}( {\overline{ B_{d}^{0}}} \rightarrow \pi^+ K^- ) \over
{\rm BR}( B_{d}^0 \rightarrow \pi^- K^+ )+ {\rm BR} ( \overline{ B_{d}^0}
\rightarrow
\pi^+ K^- )},
\nn \\     
{\cal A}_{\rm CP}^{0 0}&=&{{\rm BR}( B_{d}^0 \rightarrow \pi^0 K^0 )-
{\rm BR}(\overline{ B_{d}^0} \rightarrow \pi^0 \overline{ K^0} ) \over
{\rm BR}( B_{d}^0 \rightarrow \pi^0 K^0 )+ {\rm BR}( \overline{B_{d}^0}
\rightarrow \pi^0 \overline{
K^0 }
)}. 
\end{eqnarray}

Moreover, in the future (possibly at LHCb \cite{ball}) we will have at 
our disposal the time dependent CP-asymmetries of the neutral decay 
modes \cite{fl2} that will provide us 
with
 additional information.
Anyway, measuring all the asymmetries of eq.(\ref{eqabis}) is already a 
very 
challenging task and
 the experimental
results should be considered very
preliminar.

Our aim in this letter is to  show  how isospin symmetry allows us to
obtain relations or sum rules between 
CP-averaged branching ratios of  different modes and also between  
CP-asymmetries. 
Some of these relations were known in the limit of no isospin breaking 
\cite{lip},
some in the context of the SM \cite{n1,gron} and some of them are new. 
These sum rules will be derived
in a transparent way and they will be valid for any model.
 They will be useful to understand in a more direct way the 
implications of the present
experimental results in terms of isospin breaking.
From our estimate of how reasonably large could be
isospin breaking we can  try to 
`test' the experimental results.
We will show, in that respect, that present 
experimental results
are quite unexpected. We will first
construct these sum rules in a model independent way, then we will 
analyze the case of the Standard Model
  in
the framework of
NLO QCD factorization \cite{bbns00}
and, in the last section, we will test the sum rules using present 
experimental data.   
\bigskip

\bigskip
\centerline{\bf I. Isospin decomposition} 
\bigskip

In the theoretical description of the decay modes of $B \rightarrow \pi 
K$, isospin symmetry and its
breaking plays
a central role. Indeed, the CP-averaged branching ratios $R$, $R_c$ and 
$R_0$ can be 
considered as a measure of 
isospin breaking, i.e., if isospin were an exact symmetry they 
would
be
equal to one.

A general amplitude for a hadronic $B$ decay  based 
on the quark transition level ${\bar b}
\rightarrow
{\bar s} q {\bar q}$  is described by an effective lagrangian 
\cite{bbl} that 
includes current-current
operators, QCD  and electroweak penguins.                

Using isospin decomposition one arrives easily to the following 
relations: 
\begin{eqnarray} \label{eqc}
-\sqrt{2} A \left( B^+ \rightarrow \pi^0 K^+ \right)&=&
A \left( B^+ \rightarrow \pi^+ K^0 \right) + d_1,
\nn \\
-A \left( B^0 \rightarrow \pi^- K^+ \right)&=&
A \left( B^+ \rightarrow \pi^+ K^0 \right) + d_2, \nn \\
\sqrt{2} A \left( B^0 \rightarrow \pi^0 K^0 \right)&=& A \left( B^+
 \rightarrow \pi^+ K^0 \right)
 + d_2 - d_1,
\end{eqnarray}      
where $d_{1}$ and $d_{2}$   vanish in  absence of
isospin breaking.
In  presence of New Physics the most general expression for $d_1$ and
$d_2$ is\footnote{Latin subindex $i$ will always be understood to run 
from 1 to 2.}

\begin{equation} \label{di}
d_{i}=|P| \xi_i e^{i \theta_i } \left( e^{i \gamma} - a_i e^{i
\phi_{a_{i}} }
-i b_i e^{i \phi_{b_{i}}} \right),
\end{equation}
where $P$ contain all CP conserving terms of the  penguin contribution
to $B^+ \rightarrow \pi^+ K^0$.
$\xi_{i}$ parametrize  isospin breaking  and they are 
expected 
to be small parameters.
$\theta_i$,
$\phi_{a_{i}}$, $\phi_{b_{i}}$ are strong phases
and  $\gamma$ and $ib_{i}$ parametrize  weak phases  that change
 sign under a CP
transformation. We will follow as close
as possible the notation of \cite{neu3}. We show explicitly in  
eq.(\ref{di}) the dependence on 
$\gamma$, meaning that $b_{1}$ and $b_{2}$ can be non-zero only if 
there is New 
Physics. 
In a similar way, we can parametrize, in general, the amplitude:
\begin{equation} \label{amp}
A \left( B^+ \rightarrow \pi^+ K^0 \right)=|P| e^{i \theta_0} \left(1 -
i
b_{0} e^{i \phi_{b_{0}} } \right),
\end{equation} 
with $ib_{0}$ changing sign under CP and $\theta_0$, $\phi_{b_{0}}$
are strong phases.  

The Standard Model limit of these parameters can be found in  
Appendix A.

If we now use eq.(\ref{eqc}) to compute eq.(\ref{eqa}) we obtain very simple
expressions that show in a
manifest way  
the relations between $R$, $R_{c}$ and $R_{0}$:
\begin{eqnarray} \label{ere1}
R&=&1+u_{+}, 
\\ \label{ere2}
R_{c}&=&1+ z_{+}, 
\\ \label{ere3}
R_0 &=&1 + n_+ = 1+ {u_+ } - z_{+} + k_1.
\end{eqnarray}
In a similar way, after choosing certain combinations of observables,  
we find the corresponding decomposition for  the CP-asymmetries:
\begin{eqnarray} \label{omega1}
{\cal A}^{-+}_{\rm CP} R &=&{\cal A}^{+ 0}_{\rm CP} + u_{-}, 
\\ \label{omega2}
{\cal A}^{0 +}_{\rm CP} R_c &=&{\cal A}^{+ 0}_{\rm CP} + z_{-},
\\ \label{omega3}
{\cal A}^{00}_{\rm CP} R_0 &=&{\cal A}^{+ 0}_{\rm CP}+n_- =
{\cal A}^{+ 0}_{\rm CP} + u_{-} -
z_{-} + k_2.
\end{eqnarray}
The exact dependence of $u_{\pm}$, $z_{\pm}$ and $k_{1}$, $k_{2}$ in 
terms of $d_{1}$, $d_{2}$ can be 
found in 
Appendix B. For the following discussion  we will only need to  
notice that $u_{\pm}$ and
$z_{\pm}$ contain a piece
linear in $\xi_i$ while  $k_1$ and $k_2$ are 
quadratic in $\xi_i$. Consequently, being $\xi_i$ a measure of 
isospin breaking, one would expect $k_{1}$ and $k_{2}$ to be 
smaller than
$u_{\pm}$ and $z_{\pm}$. 
\bigskip

\bigskip
\centerline{\bf II. Sum rules }
\bigskip

Eqs.(\ref{ere1}-\ref{ere3}) and eqs.(\ref{omega1}-\ref{omega3}) 
provide us
with all  necessary ingredients
to construct a set of `sum
rules'.
Notice that we
pretend to
find
those relations  between  CP-averaged observables ($R$) and 
CP-asymmetries (${\cal A_{\rm CP}}$) that 
minimize the
impact of isospin
breaking, i.e., we should get rid of all terms linear in $\xi_i$.

The first relation is  linear  and it is obtained by substituting 
$u_{+}$ and $z_{+}$ in
eq.(\ref{ere3}) by $R$ and 
$R_c$ 
\begin{equation} \label{i}
{\rm I}) \;\; R_{0} - R + R_c -1 = k_1.
\end{equation}

Some remarks, concerning eq.(\ref{i}), are in order here. First, this 
is an 
exact relation and it
involves
terms of order $\xi_i^2$ and higher that
measure the amount of 
isospin breaking. Second, 
$k_1$ can be interpreted, looking at eq.(\ref{defk1}) of Appendix A, as 
a 
measure of the misalignment
 between the isospin breaking
contributions to two channels: 
$\sqrt{2} A \left( B^+ \rightarrow \pi^0 K^+ \right)$ and 
$A \left( B^0 \rightarrow \pi^- K^+ \right)$. 
Even in  presence of isospin breaking 
if the new contributions to these channels are equal, 
i.e., $d_1=d_2$ then $k_1$
would be
exactly zero. On the contrary, if the isospin contribution to these 
channels have opposite sign
then $k_1$ would be maximal. We should look at data to discern 
 which of the two scenarios is closer to the one
realized in nature.

Eq.(\ref{i}) was first obtained in \cite{lip}, but with $k_1$ equal to 
zero  
and in the 
 SM case in \cite{n1}. Here, we provide  a model independent 
expression 
for  $k_1$:
\begin{eqnarray} \label{k1gen}
k_1 &=&
 {-1 \over 1 +b_{0}^2} \;\left\{\;
\xi_{1} \xi_{2} \; {\rm Re} \left[ e^{i (\theta_1 -\theta_2)
}
 \left( e^{i \gamma} - a_1 e^{i
\phi_{a_{1}} }
-i b_1 e^{i \phi_{b_{1}}} \right)  \left( e^{-i \gamma} - a_2 e^{-i
\phi_{a_{2}} }
+i b_2 e^{-i \phi_{b_{2}}} \right) \right] \right. \nn
\\&-& \left.
\xi_{1}^2 \; \left[
(1+a_{1}^2+b_{1}^2)- 2 a_{1} \;{\rm cos} \phi_{a_{1}} \;{\rm cos}
\gamma -
2 b_{1} \; {\rm cos} \phi_{b_{1}} \; {\rm sin} \gamma \right] \right\}+
 \left\{\begin{array}{ll}
 \; { \gamma}  \rightarrow {-\gamma}
\\
\; {b_{1,2} \rightarrow -b_{1,2}}
\end{array} \right.        
\end{eqnarray}                                        
We have checked, explicitly, that taking the SM limit (Appendix A) 
of eq. (\ref{k1gen}) and 
the strong phase $ \omega \rightarrow 0$, eq.(\ref{k1gen}) reduces to 
the expression 
found in 
\cite{n1}.

Following exactly the same steps, substituting $u_-$ and $z_-$ in 
eq.(\ref{omega3}), we obtain
the second sum rule in  a quite transparent way:
\begin{equation} 
{\rm II}) \;\; {\cal A}^{00}_{\rm CP} R_0 - 
{\cal A}^{-+}_{\rm CP} R+ {\cal A}^{0 +}_{\rm CP}
R_c
-{\cal A}^{+ 0}_{\rm CP} = k_2.
\end{equation}      

This is also an exact relation and it was found in \cite{n1}
in the SM case. We can also 
interpret $k_2$ from
eq.(\ref{defk2}) 
as a measure of the misalignment between 
$d_1$ and $d_2$, then  the same conditions that force $k_1$ to vanish 
also apply to $k_2$.
But, more specifically,  $k_2$ measures the importance of weak phase 
differences between
$d_1$ and its CP conjugate and between $d_2$ and its CP conjugate, i.e.,
 $k_2=0$ if
$d_1={\overline d_1}$ and $d_2={\overline d_2}$.
\begin{figure}[t]
\vspace*{-1cm}
$$\hspace*{-1.cm}
\epsfysize=0.3\textheight
\epsfxsize=0.3\textheight
\epsffile{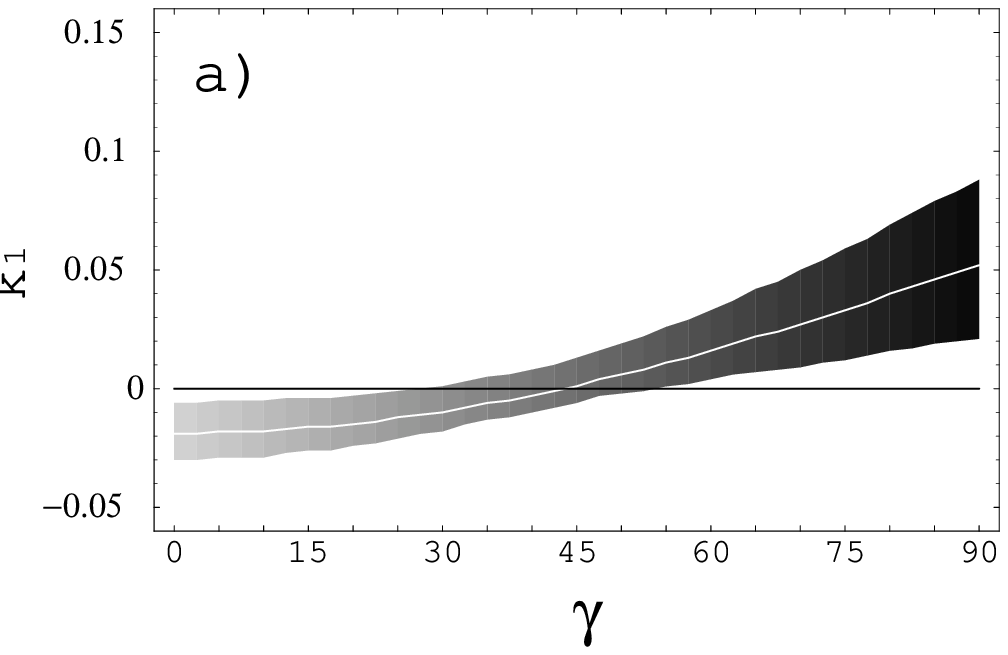} \hspace*{0.3cm}
\epsfysize=0.3\textheight
\epsfxsize=0.3\textheight
 \epsffile{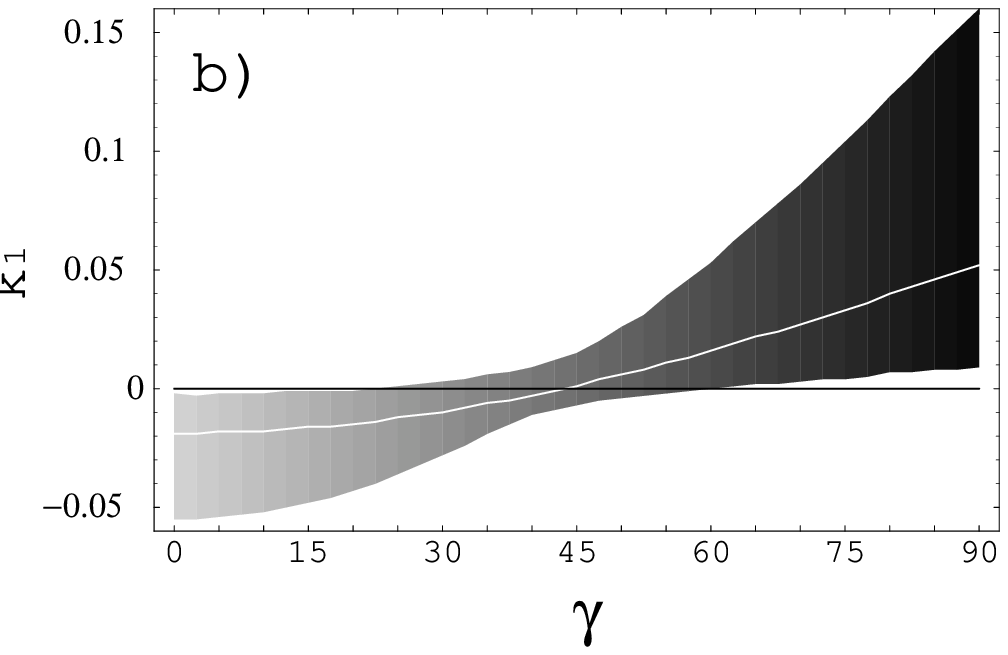}
$$
\vspace*{-0.9cm}
\caption[]{Sum rule {\rm I} evaluated for the SM  using 
NLO QCD 
factorization \cite{bbns00} for  
values of $\gamma$ in the first 
quadrant: (a) low uncertainty ($\varrho_A=1$) from annihilation 
topologies,
(b) large uncertainty ($\varrho_A=2$) from annihilation topologies.
}\label{fig:fig1}
$$\hspace*{-1.cm}      
\epsfysize=0.3\textheight
\epsfxsize=0.3\textheight
\epsffile{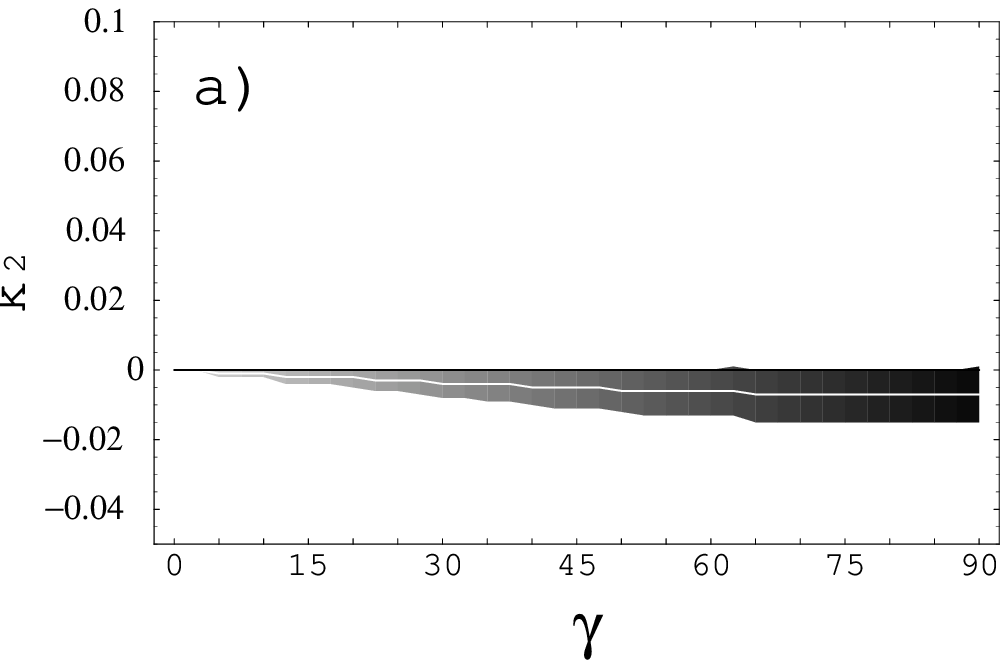} \hspace*{0.3cm}
\epsfysize=0.3\textheight
\epsfxsize=0.3\textheight
 \epsffile{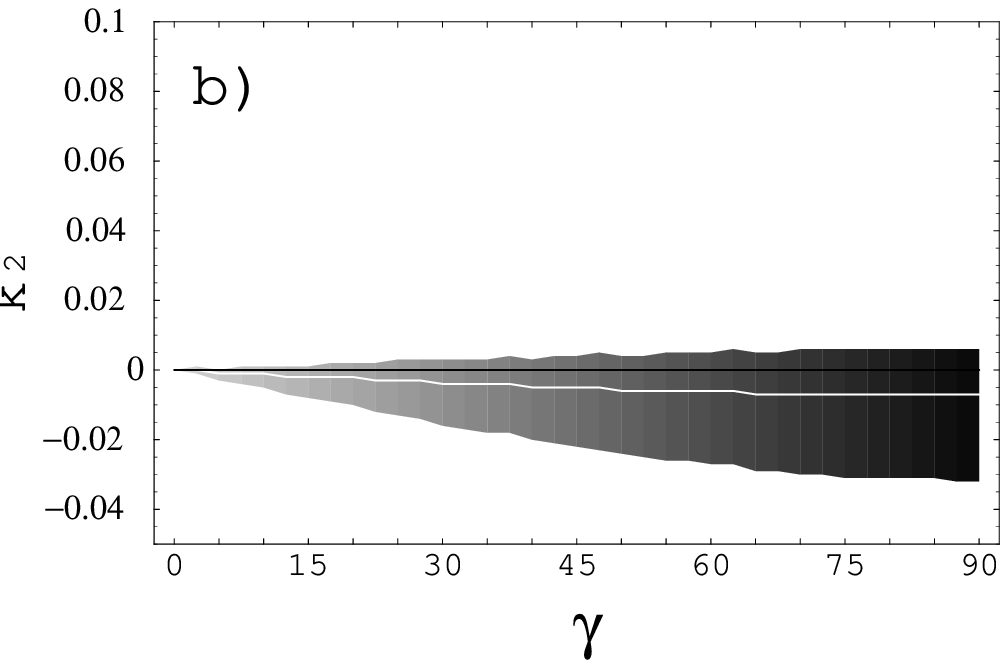}
$$
\vspace*{-0.9cm}
\caption[]{Sum rule II with same conditions as in Fig.\ref{fig:fig1}
}\label{fig:fig2}
\end{figure}
The model independent expression for $k_2$ is:
\begin{eqnarray} \label{k2gen}
k_2 &=& {-1 \over 1 +b_{0}^2} \;\left\{\;
\xi_{1} \xi_{2} \; {\rm Re} \left[ \; e^{i (\theta_1 -\theta_2) }
 \left( e^{i \gamma} - a_1 e^{i
\phi_{a_{1}} }
-i b_1 e^{i \phi_{b_{1}}} \right)  \left( e^{-i \gamma} - a_2 e^{-i
\phi_{a_{2}} }
+i b_2 e^{-i \phi_{b_{2}}} \right) \right]  \right. \nn \\
 &+& \left.  2\; \xi_{1}^2 \; \left[
a_{1} \;{\rm sin} \phi_{a_{1}} \;{\rm sin} \gamma -
 b_{1} \; {\rm sin} \phi_{b_{1}} \; {\rm cos} \gamma
- a_{1} b_{1} \; {\rm sin}(\phi_{a_{1}}-\phi_{b_{1}})
\right] \right\}- \left\{\begin{array}{ll}
 \; { \gamma}  \rightarrow {-\gamma}
\\
\; {b_{1,2} \rightarrow -b_{1,2}}
\end{array} \right.      
\end{eqnarray}

 We have also checked  that its SM limit agrees with
\cite{n1}.
In Fig.\ref{fig:fig1} and \ref{fig:fig2} we have plotted sum rule I and 
II versus $\gamma$ 
using the predictions for the Standard Model in the framework of NLO 
QCD   
factorization \cite{bbns00}. We have restricted ourselves
to the first quadrant for $\gamma$ since it is the region where the 
standard Unitarity Triangle Analysis (UTA)
expects $\gamma$ to lie. Indeed the precise value is 
\cite{silv,ciu}
$$ \label{gamma}
\gamma=\left(54.8 \pm 6.2 \right)^0.
$$

\begin{figure}[p]
\vspace*{-1cm}
$$\hspace*{-1.cm}
\epsfysize=0.3\textheight
\epsfxsize=0.3\textheight
\epsffile{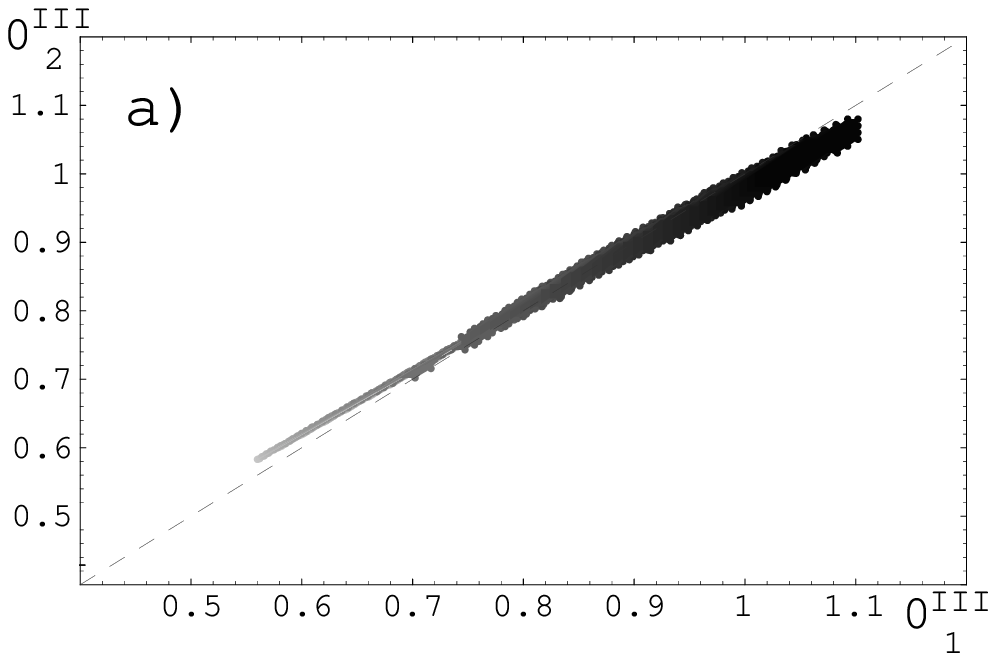} \hspace*{0.3cm}
\epsfysize=0.3\textheight
\epsfxsize=0.3\textheight
 \epsffile{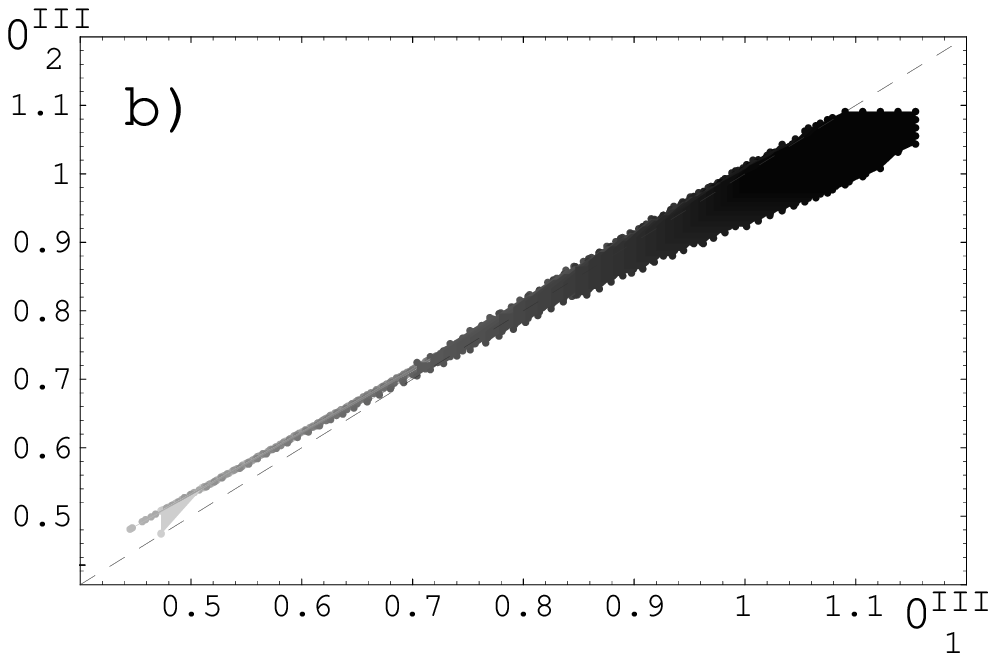}
$$
\vspace*{-1.2cm}
\caption[]{Correlation between $O_1^{\rm III}$ and $O_2^{\rm III}$ 
evaluated in the SM using NLO QCD
factorization, for low (a) and large (b) uncertainty coming 
from annihilation 
topologies. The lighter region corresponds to the lowest values of 
$\gamma$ 
inside the first quadrant.
}\label{fig:fig3}
$$\hspace*{-1.cm}
\epsfysize=0.3\textheight
\epsfxsize=0.3\textheight
\epsffile{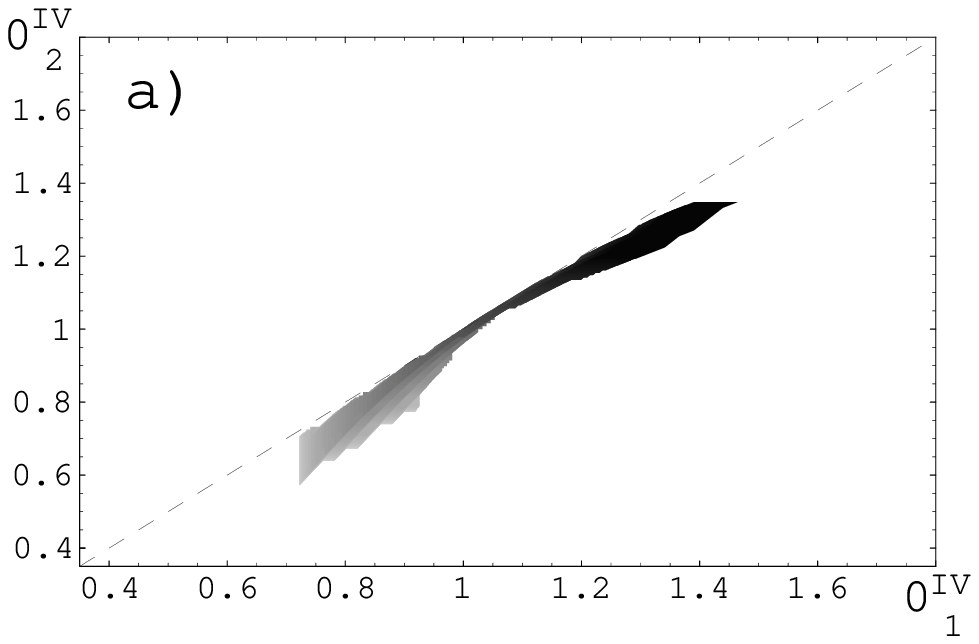} \hspace*{0.3cm}
\epsfysize=0.3\textheight
\epsfxsize=0.3\textheight
 \epsffile{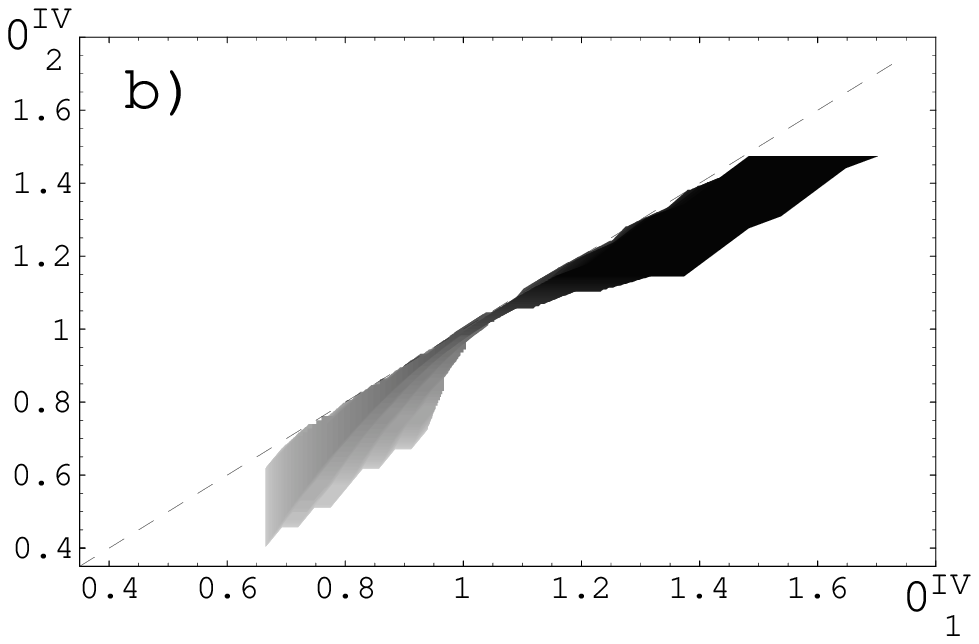}
$$
\vspace*{-1.2cm}
\caption[]{Correlation between $O_1^{\rm IV}$ and $O_2^{\rm IV}$ 
of sum rule {\rm IV}, conventions as in Fig.\ref{fig:fig3}.
}\label{fig:fig4}
$$\hspace*{-1.cm}                      
\epsfysize=0.3\textheight
\epsfxsize=0.3\textheight
\epsffile{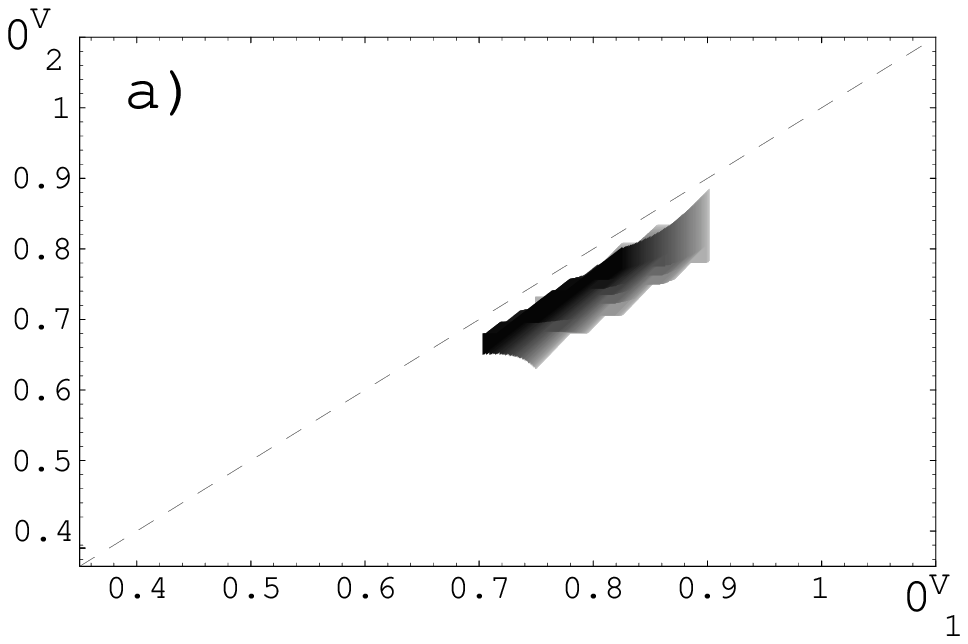} \hspace*{0.3cm}
\epsfysize=0.3\textheight
\epsfxsize=0.3\textheight
 \epsffile{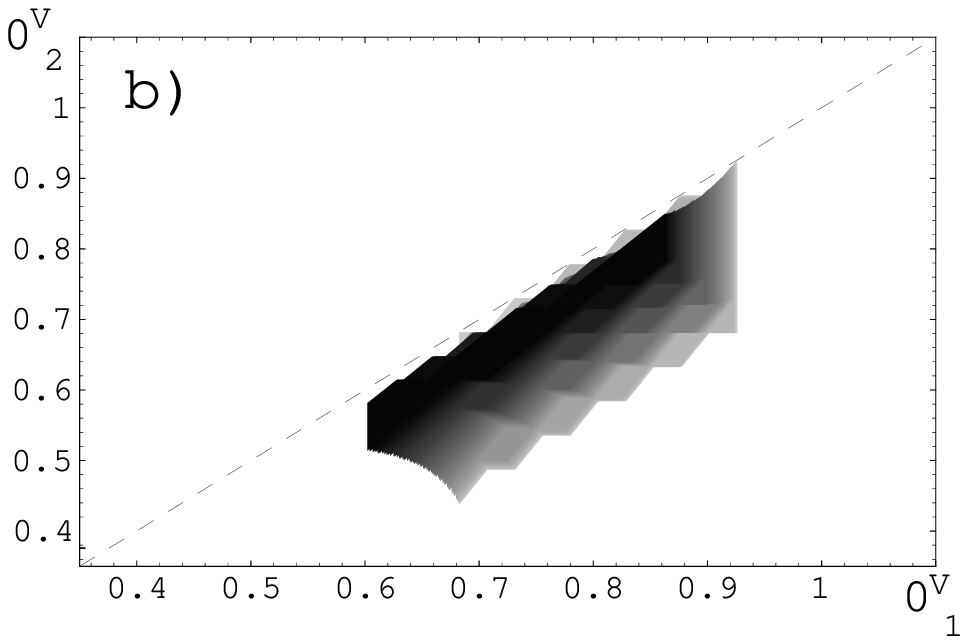}
$$
\vspace*{-1.2cm}
\caption[]{Correlation between $O_1^{\rm V}$ and $O_2^{\rm V}$ of sum 
rule {\rm V}, conventions as in Fig.\ref{fig:fig3}.
}\label{fig:fig5}
\end{figure}

\begin{figure}[p]
\vspace*{-1cm}
$$\hspace*{-1.cm}
\epsfysize=0.3\textheight
\epsfxsize=0.3\textheight
\epsffile{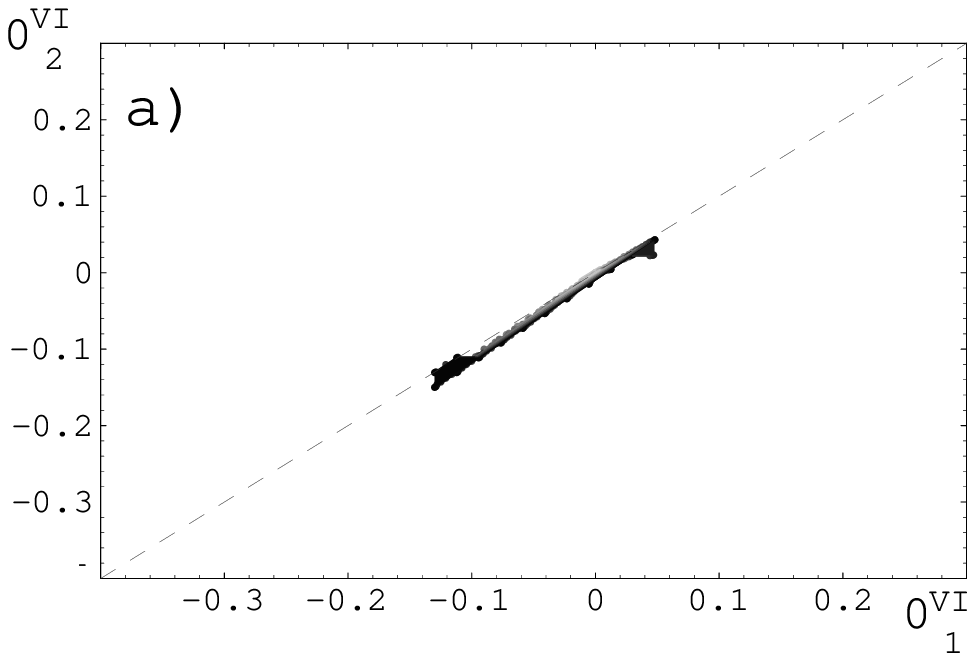} \hspace*{0.3cm}
\epsfysize=0.3\textheight
\epsfxsize=0.3\textheight
 \epsffile{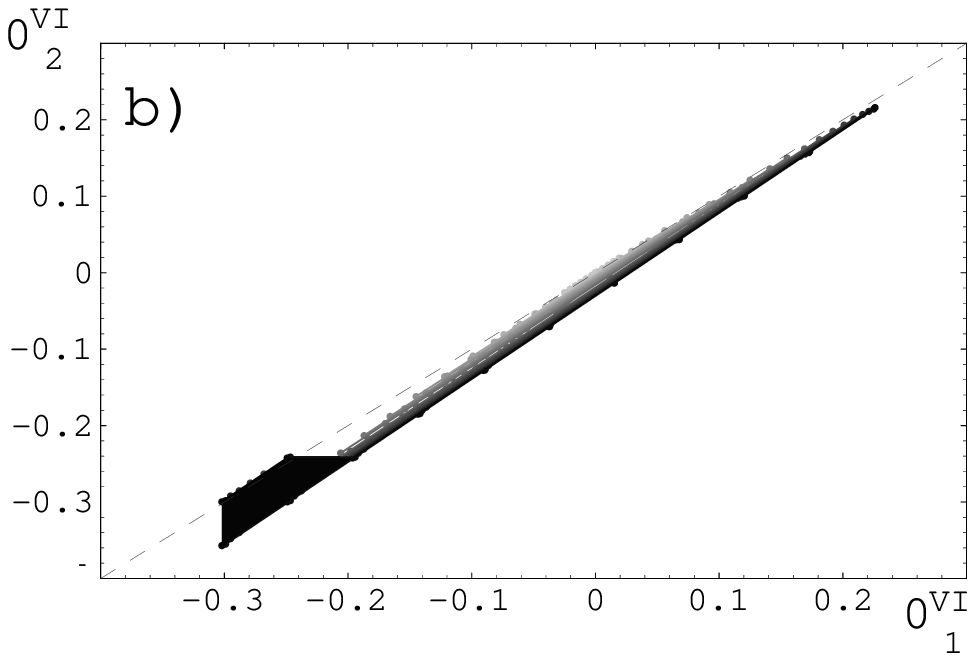}
$$
\vspace*{-1.2cm}         
\caption[]{Correlation between $O_1^{\rm VI}$ and 
$O_2^{\rm VI}$, conventions as in Fig.\ref{fig:fig3}.    
}\label{fig:fig6}
$$\hspace*{-1.cm}
\epsfysize=0.3\textheight
\epsfxsize=0.3\textheight
\epsffile{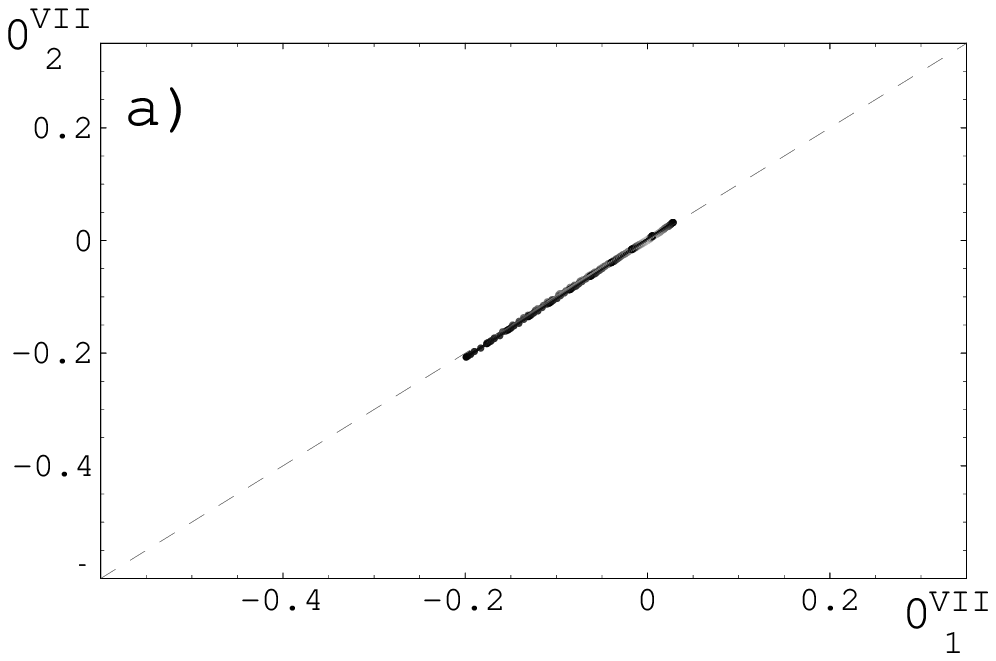} \hspace*{0.3cm}
\epsfysize=0.3\textheight
\epsfxsize=0.3\textheight
 \epsffile{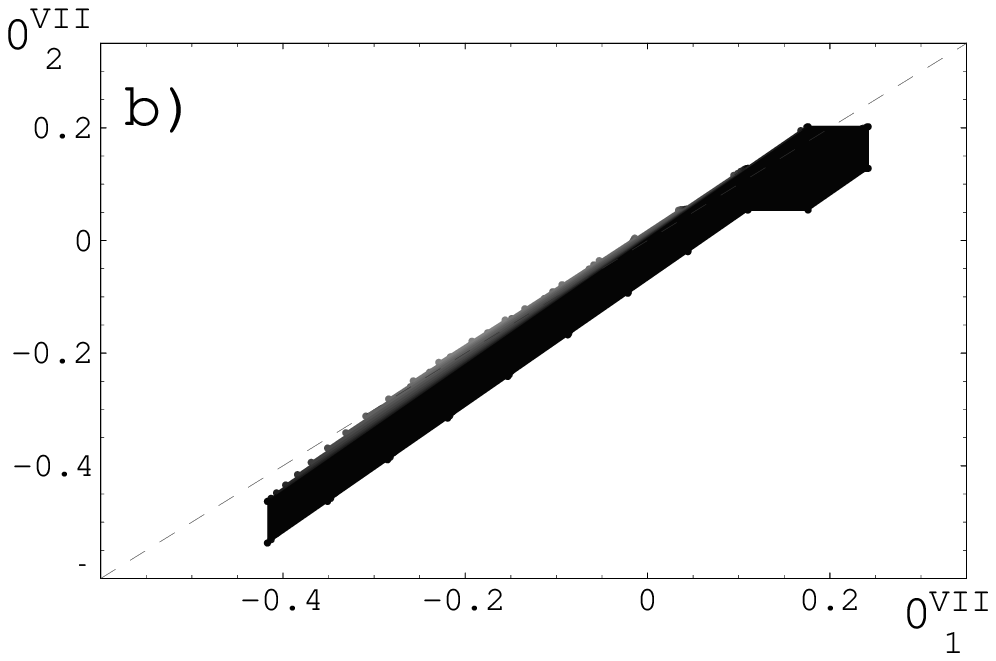}
$$
\vspace*{-1.2cm}
\caption[]{Correlation between 
$O_1^{\rm VII}$ and $O_2^{\rm VII}$, conventions as in Fig.\ref{fig:fig3}.    
}\label{fig:fig7}
$$\hspace*{-1.cm}
\epsfysize=0.3\textheight
\epsfxsize=0.3\textheight
\epsffile{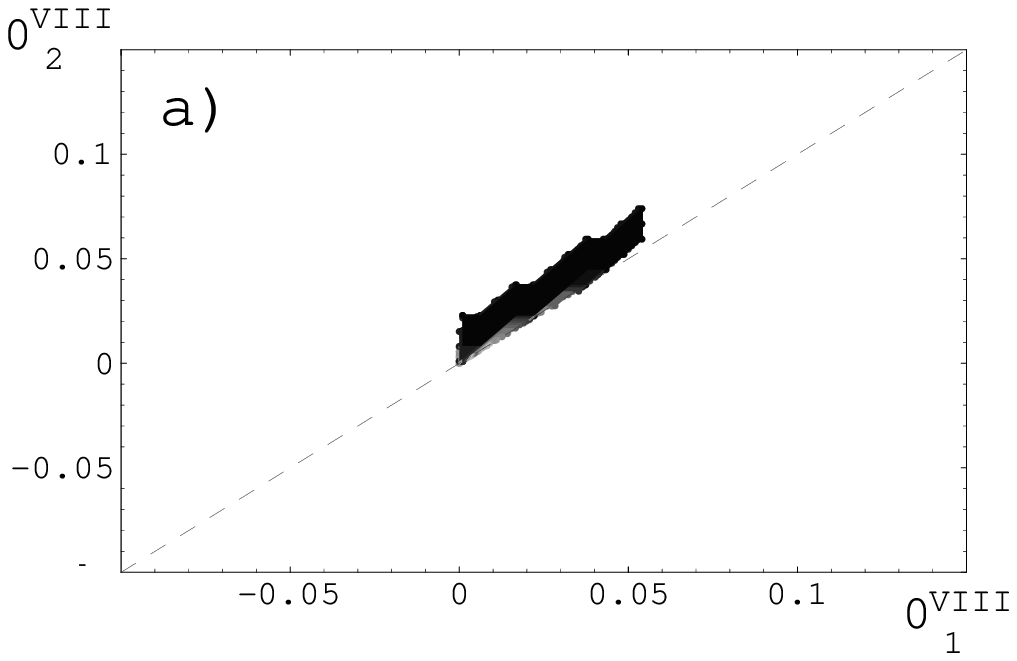} \hspace*{0.3cm}
\epsfysize=0.3\textheight
\epsfxsize=0.3\textheight
 \epsffile{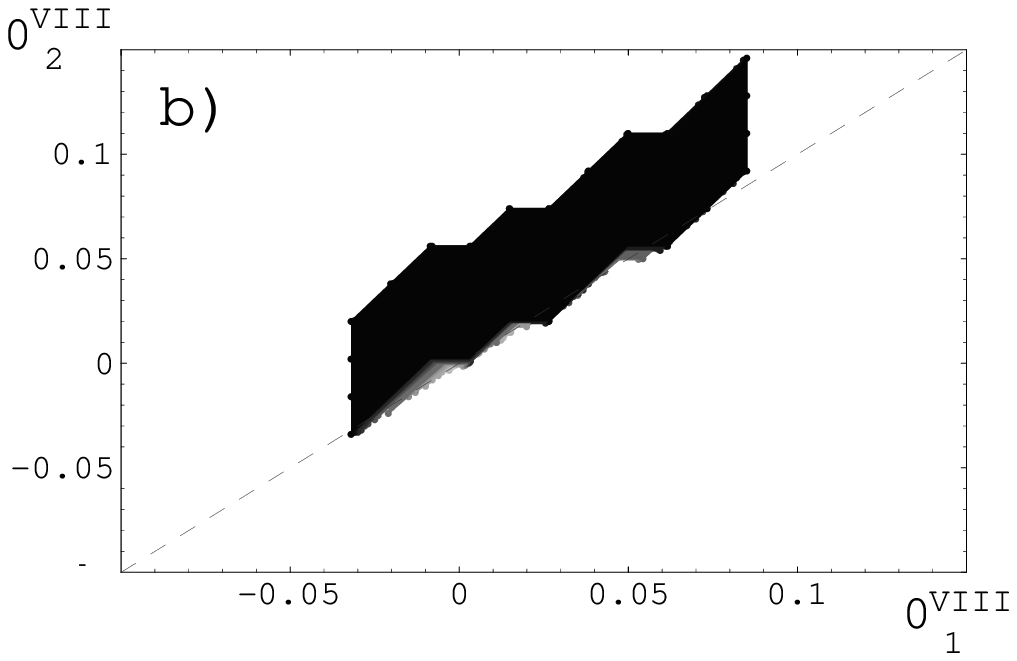}
$$
\vspace*{-1.2cm}
\caption[]{Correlation between $O_1^{\rm VIII}$ and 
$O_2^{\rm VIII}$  conventions as in Fig.\ref{fig:fig3}.    
}\label{fig:fig8}
\end{figure}

We have plotted the result taking into account two cases for the 
estimate of the uncertainty coming 
from the annihilation topologies (see  \cite{bbns00}): low 
uncertainty (a) 
and extremely large uncertainty (b). 
The low uncertainty case corresponds to the more realistic situation 
of taking the
parameter $\varrho_A$ (defined in \cite{bbns00}) equal to one and the 
large uncertainty
case corresponds to the extreme case (very conservative) of  
$\varrho_A$ 
equal to two.
The gradient from light to dark it 
is used for further reference 
in Figs.\ref{fig:fig3}-\ref{fig:fig9} to indicate the different values 
of 
$\gamma$.  
Notice that around the UTA value of  $\gamma$ \ref{gamma} the values of 
$k_1$ and $k_2$ are extremely 
small: $-0.01 \lsim k_1 \lsim 0.05 $  and $-0.030 \lsim k_2 
\lsim 0.005$. 

\begin{table}[t]
\caption{Strongly correlated observables associated to sum rules 
{\rm III-V}}
\label{tab:tab1}
\vspace*{0.2cm}
\begin{center}
\begin{tabular}{|lll|}\hline
\rule[-0.3cm]{0cm}{0.9cm}
${\rm III}$
 & ${\cal O}^{\rm III}_1=R$ & ${\cal O}^{\rm III}_2=R_0 R_c$  \\
\hline
\rule[-0.3cm]{0cm}{0.9cm}
${\rm IV}$
& ${\cal O}^{\rm IV}_1=R_c$ & ${\cal O}^{\rm IV}_2=-R_0/R + 2$ \\
  \hline
\rule[-0.3cm]{0cm}{0.9cm}
${\rm V}$
 & ${\cal O}^{\rm V}_1=R_0$ & ${\cal O}^{\rm V}_2=-R_c/R + 2$ \\
 \hline
\end{tabular}
\end{center}
\end{table}                

The following set of sum rules are obtained in a completely different 
way. The aim is to find the simplest combinations of CP-averaged 
branching ratios ($R$) that would be
strongly correlated if isospin were an exact symmetry. 
They are obtained combining  eq.(\ref{ere1}-\ref{ere3}) to construct 
a quantity of order $\xi_i^2$, dividing by one of the $R$'s ($R$, $R_c$ 
or 
$R_0$) and 
reinserting 
sum rule I to simplify the expression.
The result is the following three sum rules:
\begin{equation}
{\rm III}) 
\;\; R = {R_0 R_c}  +k_3,        
\end{equation}   
with $
k_3= z_+ \left(z_+-u_+\right)-k_1-k_1 z_+.
$     
\begin{equation}
{\rm IV}) 
\;\; R_c = -{R_0 \over R} + 2 +k_4,       
\end{equation}     
with     
$
k_4= {(u_+ z_+ + k_1) / (1 + u_+)}.
$
And, finally,

\begin{equation}
{\rm V}) 
\;\; R_0 = -{R_c \over R} + 2 +k_5.      
\end{equation}         
with $
k_5=k_1 + {u_+ (u_+ - z_+) /(1 + u_+)}$.
This  sum rule can be  related  with one proposed in \cite{n1} 
(but with the inverse $R/R_c$) for 
the SM
case and in an
approximate
form, i.e., keeping only the term $\xi^2_i$.

Notice that the expressions for $k_3$, $k_4$ and $k_5$ are model 
independent and their general expressions are obtained using the
expressions of $u_+$, $z_+$ in Appendix B and $k_1$  of 
eq.(\ref{k1gen}).

Sum rules {\rm III} to {\rm V} will allow us to  define a set of 
observables. These observables are chosen in such
 a way to be strongly correlated by isospin, 
i.e., they should have the same 
value except for corrections of order $\xi_i^2$. They are given in 
Table~\ref{tab:tab1}. 
A nice graphical interpretation of these sum rules can be obtained 
by plotting
 ${\cal O}_1^\alpha$ versus ${\cal 
O}_2^\alpha$ ($\alpha={\rm III,IV,V}$) for different values of 
$\gamma$.
In Figs.\ref{fig:fig3}-\ref{fig:fig5} we  shown the prediction for 
these observables for the SM using
NLO QCD factorization  \cite{bbns00}   taking into account 
the uncertainty coming from  
annihilation topologies (low (a) and high (b)).
The region presented in Figs.\ref{fig:fig3}-\ref{fig:fig5} corresponds 
to 
varying all 
the
parameters (amplitudes and strong phases) within
the predictions of NLO QCD factorization for the SM.
All figures are restricted to values of $\gamma$ ($0 \leq \gamma 
\leq \pi/2$)
 inside the first quadrant, according
to the SM fit
from other measurements. Shading should be understood in the 
following way: lighter region corresponds 
exclusively to
low values of $\gamma$, following the pattern of 
Fig.\ref{fig:fig1}-\ref{fig:fig2}, while the 
dark
region can correspond to large or small values of $\gamma$ inside the 
first quadrant, since they cannot be distinguished in the plots.
 
In absence of isospin breaking 
 both observables should fall  in the diagonal of 
Figs.\ref{fig:fig1}-\ref{fig:fig3}, 
with ${\cal O}_i^{\alpha}=1$. If
isospin breaking is small $O_1^{\alpha}$ and $O_2^{\alpha}$ should stay
near the diagonal.       
The deviation from one
 {\bf along the diagonal} gives an idea of the 
isospin 
breaking terms of order $\xi_i$ (remember that $R$, $R_c$ and $R_0$ 
 measure isospin breaking of this 
size). This is useful to have an idea of the maximal size of this breaking.
Notice that it also implies that each pair of observables 
($O_1^{\alpha}$, $O_2^{\alpha}$) is chosen in 
such a way to present the same deviation of order $\xi_i$, 
independently 
of the model.

More interestingly, deviations {\bf from the diagonal} 
 would
measure isospin breaking
contributions of order  $\xi_i^2$. 
It implies 
that if isospin  is not badly broken, we can estimate that the
 deviations from the diagonal will be smaller 
than the square of the maximal deviation from one along the diagonal. For
 instance, in Fig.\ref{fig:fig4}b the 
maximal deviation from one along the diagonal is approximately $0.5$, 
then the 
maximal expected deviation from the 
diagonal would be 0.25 and, indeed, this is the case. 
This rule applies to all figures evaluated using NLO QCD factorization.

An experimental measurement of these observables very far away from 
the diagonal and the SM allowed region 
would require the 
contribution of isospin breaking New Physics.
Present data favors this situation as we will see in the next section.

\begin{table}[t]
\caption{Strongly correlated observables associated to sum rules 
{\rm VI-VIII}}
\label{tab:tab2}
\vspace*{0.2cm}
\begin{center}
\begin{tabular}{|lll|}\hline
\rule[-0.3cm]{0cm}{0.9cm}
$\hspace*{-0.2cm}{\rm VI}$
& $\hspace*{-0.25cm}{\cal O}^{\rm VI}_1={\cal A}_{\rm CP}^{-+} R$ & 
$\hspace*{-0.2cm}{\cal O}^{\rm 
VI}_2=A_{\rm
CP}^{+0}-1+\left(1+ {\cal A}_{\rm CP}^{00} R_0-{\cal A}_{\rm CP}^{+0}  
\right)\left(
1+ {\cal A}_{\rm CP}^{0+} R_c - {\cal A}_{\rm CP}^{+0} \right)$ \\
\hline
\rule[-0.3cm]{0cm}{0.9cm}
$\hspace*{-0.2cm}{\rm VII}$
& $\hspace*{-0.25cm}{\cal O}^{\rm VII}_1={\cal A}_{\rm CP}^{0+} R_c$ &  
  $\hspace*{-0.2cm}{\cal 
O}^{\rm VII}_2=A_{\rm
CP}^{+0}+\left({\cal A}_{\rm CP}^{-+} R-{\cal A}_{\rm CP}^{00} R_0 
\right)/\left(
1+{\cal A}_{\rm CP}^{-+} R - {\cal A}_{\rm CP}^{+0} \right)$ \\
  \hline                         
\rule[-0.3cm]{0cm}{0.9cm}
$\hspace*{-0.2cm}{\rm VIII}$
& $\hspace*{-0.25cm}{\cal O}^{\rm VIII}_1={\cal A}_{\rm CP}^{00} R_0$ &  
 $\hspace*{-0.2cm}{\cal 
O}^{\rm VIII}_2=A_{\rm
CP}^{+0}+\left({\cal A}_{\rm CP}^{-+} R-{\cal A}_{\rm CP}^{0+} R_c 
\right)/\left(
1+{\cal A}_{\rm CP}^{-+} R - {\cal A}_{\rm CP}^{+0} \right)$ \\
 \hline
\end{tabular}
\end{center}
\end{table}         

Finally, following the same strategy as in sum rule III-V, we can 
construct a set of three sum rules 
involving the CP-asymmetries using the  translation table
\begin{eqnarray}
R &\rightarrow& {\cal A}_{\rm CP}^{-+} R-{\cal A}_{\rm CP}^{+0}+1, \nn 
\\
R_c &\rightarrow& {\cal A}_{\rm CP}^{0+} R_c-{\cal A}_{\rm CP}^{+0}+1, 
\nn \\   
R_0 &\rightarrow& {\cal A}_{\rm CP}^{00} R_0-{\cal A}_{\rm CP}^{+0}+1, 
\end{eqnarray}
and rearranging the result in the following way:
\begin{equation}
{\rm VI})
\;\; {\cal A}_{\rm CP}^{-+} R = {\cal A}_{\rm CP}^{+0} -1 + 
\left(1+ {\cal A}_{\rm CP}^{00} R_0-{\cal A}_{\rm CP}^{+0}  
\right)\left(
1+ {\cal A}_{\rm CP}^{0+} R_c - {\cal A}_{\rm CP}^{+0} \right)+k_6.
\end{equation}          
Here $k_6$  is given by $
k_6= z_- \left(z_--u_-\right)-k_2-k_2 z_-      
$.      

\begin{equation}
{\rm VII})
\;\; {\cal A}_{\rm CP}^{0+} R_c = {\cal A}_{\rm CP}^{+0} + {A_{\rm 
CP}^{-+} R 
- 
{\cal A}_{\rm CP}^{00} R_0 \over 1 + {\cal A}_{\rm CP}^{-+} R-A_{\rm 
CP}^{+0}}+k_7, 
\end{equation}
with 
$
k_7= {(u_- z_- + k_2) / (1 + u_-)}.
$     
And finally, the equivalent to sum rule V for the CP-asymmetries would 
be,
\begin{equation}
{\rm VIII})
\;\; {\cal A}_{\rm CP}^{00} R_0 = {\cal A}_{\rm CP}^{+0} + {A_{\rm 
CP}^{-+} R 
-
{\cal A}_{\rm CP}^{0+} R_c \over 1 + {\cal A}_{\rm CP}^{-+} R-A_{\rm 
CP}^{+0}}+k_8.
\end{equation}
Here $k_8$  is given by,
$$
k_8=k_2 + {u_- \left(u_- - z_-\right) \over 1 + u_-}. 
$$

The corresponding set of correlated observables associated to these sum 
rules is shown in Table \ref{tab:tab2} and their SM predictions using 
NLO QCD factorization are presented in 
Figs.\ref{fig:fig6}-\ref{fig:fig8}. It is 
remarkable 
the extreme correlation of some observables like the pair $O_1^{\rm 
VI}-O_2^{\rm VI}$ or $O_1^{\rm VII}-O_2^{\rm VII}$ .
Notice that  sum rules {\rm VI} to {\rm VIII} are chosen in such a way 
that can be
combined easily with sum rules {\rm III} to {\rm V} to eliminate $R$, 
$R_c$ or
$R_0$ keeping the dependence on $k_1$ and $k_2$. For example, 
combining sum rule {\rm III} with sum rule {\rm VI}, we can get
rid of $R$ or sum rule {\rm IV} with {\rm VII} eliminates $R_c$ and 
{\rm 
V} 
with {\rm VIII} cancels
$R_0$.
\bigskip

\bigskip
\centerline{\bf III. Numerical results and tests of the sum rules}
\bigskip
        
In this section we will evaluate the sum rules presented in the 
previous 
section using available experimental data \cite{exp,exp2} on 
CP-averaged 
branching 
ratios and CP-asymmetries (see Table~\ref{tab:tab3}).

We will start obtaining the basic block elements of the sum rules, 
i.e, the set of CP-averaged branching ratios\footnote{Notice that in 
eq.(\ref{eqa}) we are neglecting the small phase space difference 
between $B^{\pm} \rightarrow \pi^{\pm} K^0$ and $B_d^0 \rightarrow 
\pi^{\pm} K^{\mp}, \pi^0 K^0$.}: 
\begin{equation} \label{expr}
R=1.00 \pm 0.18, \quad \quad R_c=1.41 \pm 0.29, \quad \quad R_0=1.21 
\pm 
0.35.
\end{equation}
They imply  the following values for the deviations from unity,
\begin{equation} \label{expome}
u_+= 0.00 \pm 0.18, \quad \quad z_+= 0.41 \pm 0.29, \quad \quad  
n_+= 
0.21 \pm 
0.35.
\end{equation}

Following the construction of  sum rule I
eq.(\ref{i}) from
eqs.(\ref{ere1}-\ref{ere3}), and taking
the experimental  values for $u_+$, $z_+$ and $n_+$
eq.(\ref{expome}) we find a first
surprising result. While $u_+$, $z_+$  and $n_+$ are quantities of
order $\xi_i$, the experimental value of $k_1$, a quantity of order 
$\xi_i^2$, obtained from 
eq.(\ref{i}) is
\begin{equation} \label{k1exp}
k_1=0.62 \pm 0.45,
\end{equation}
with an unexpectedly large central value, although with a large error. 
The reason is that $z_+$ goes in the wrong direction and requires $k_1$ 
to be very large in order to compensate it and to reproduce the 
experimental result for 
$R_0$. Moreover, if we  compare this experimental result  
with the SM prediction  
using NLO QCD
factorization we find that the central value of the experimental 
result  is one order of 
magnitude larger than the prediction. However, given the large 
experimental error we are still
only one standard deviation away from 
the the SM prediction, so it is crucial that experimentalist try to 
reduce this error.

If the error is reduced and the central value does not change 
drastically, one should conclude  that $\xi_i$ is not small
enough to be
considered as a good expansion
parameter and that there is some new mechanism that generates
 a very large isospin breaking contribution. On the contrary, if 
future data prefer a  
central value smaller  by an order of magnitude with reduced errors 
then the small isospin breaking scenario of the SM will be again in 
good shape and sum rule I will be fulfilled.  It is interesting 
to notice that $z_{+}$ seems to be quite large as compared to $u_{+}$
 and both $z_+$ and $k_1$ depend on $d_1$, so the problem
seems to affect more to $d_1$ while $d_2$ is inside the 
expectations. 
 $d_1$ is related to
the
contributions of the charged channel:
 $A \left( B^+ \rightarrow \pi^0 K^+ \right)$.
If these experimental results are confirmed it seems that one should 
look for a type of isospin breaking New Physics
affecting more the charged than the neutral channel.

\begin{table}[t]
\caption{Measured branching ratios and CP-asymmetries for $B 
\rightarrow
\pi K$ modes. The branching ratios correspond to the average of the
three experiments BELLE, Babar and CLEO and are taken from \cite{silv}.
CP-asymmetries correspond to CLEO data \cite{exp2}}
\label{tab:tab3}
\vspace*{0.2cm}
\begin{center}
\begin{tabular}{|ll|}\hline
\rule[-0.3cm]{0cm}{0.9cm}
$BR(B_d^0 \rightarrow \pi^{\mp} K^{\pm})$  & $17.2
\pm 1.6$
\\ \hline
\rule[-0.3cm]{0cm}{0.9cm}
$BR(B^{\pm} \rightarrow \pi^{0} K^{\pm})$
  & $12.1 \pm 1.7$ \\
\hline
\rule[-0.3cm]{0cm}{0.9cm}
$BR(B^{\pm} \rightarrow \pi^{\pm} K^{0})$ & $17.2 \pm
2.6$ \\
\hline
\rule[-0.3cm]{0cm}{0.9cm}
$BR(B_d^0 \rightarrow \pi^{0} K^{0})$
& $10.4 \pm
2.6$ \\                         
\hline
\rule[-0.3cm]{0cm}{0.9cm}
${\cal A}^{+0}_{\rm CP}$ & $-0.18 \pm 0.24$ \\
\hline
\rule[-0.3cm]{0cm}{0.9cm}
${\cal A}^{0+}_{\rm CP}$ & $0.29 \pm 0.23$ \\
\hline
\rule[-0.3cm]{0cm}{0.9cm}
${\cal A}^{-+}_{\rm CP}$ & $0.04 \pm 0.16$ \\ \hline
\end{tabular}
\end{center}
\end{table}

Let's continue the analysis of  sum rules involving only
CP-averaged branching ratios, i.e, {\rm III} to {\rm V}.
In this case we will evaluate their associated observables and the
$\xi_i^2$ isospin breaking given by $k_{3}$, $k_{4}$ and $k_{5}$. The 
results are the
following:
\begin{eqnarray} \label{result}
&{\rm III})& \quad O_1^{\rm III}=1.00 \pm 0.18, \quad \quad O_2^{\rm 
III}=1.70 
\pm 0.71, \quad \quad k_3=-0.70 \pm 0.62. \nn \\
&{\rm IV})& \quad O_1^{\rm IV}=1.41 \pm 0.29, \quad \quad O_2^{\rm 
IV}=0.79
\pm 0.32, \quad \quad k_4=0.62 \pm 0.43. \nn \\      
&{\rm V})& \quad O_1^{\rm V}\;=1.21 \pm 0.35, \quad \quad O_2^{\rm 
V}\;=0.59
\pm 0.23, \quad \quad k_5=0.62 \pm 0.43.       
\end{eqnarray}
These results imply that, with  present experimental data,  none of 
the sum rules {\rm III} to {\rm V} seems to follow the expected 
behavior of a small $\xi_i^2$ term, i.e, 
to fall in or near 
the diagonal of Figs.~\ref{fig:fig3} to \ref{fig:fig5}.
According to the discussion of the previous section, while 
eq.(\ref{expome}) corresponds to the deviation from one along the 
diagonal and measures the isospin breaking of order $\xi_i$, 
$k_{j}$ with $j=3,4,5$ of eq.(\ref{result}) gives an idea of the 
deviation
from the diagonal (more precisely $k_{j}/\sqrt{2}$). 
We observe 
that, 
similarly to what happens with sum rule {\rm I},  
  data force  quantities that are formally of order
$\xi_i^2$ like
$k_{j}$ to be of the same size or larger than quantities 
of order $\xi_i$ like $u_+$ or $z_+$ of  
 eq.(\ref{expome}). Moreover, 
this experimental result is in conflict,
at least for the central values, with 
the predictions for the SM using NLO QCD factorization shown in 
Figs.~\ref{fig:fig3} 
to \ref{fig:fig5}.

The set of sum rules {\rm II} and {\rm VI} to {\rm VIII}  
involves the still non measured  
${\cal A}_{\rm CP}^{00}$, so it is not possible to compare data with 
the SM predictions. However, under certain assumptions
on isospin and future data we can explore what type of information they 
can give us on 
this CP-asymmetry.  

From Table \ref{tab:tab3} it is possible to evaluate
 $u_- $ and  $z_- $,
 eq.(\ref{omega1}) and eq.(\ref{omega2}):
\begin{equation} \label{wzmc}
u_{-}=  0.22 \pm 0.29,\quad \quad
z_{-}= 0.59 \pm 0.41.
\end{equation}       

We can also evaluate some of the observables of sum rule {\rm VI} to 
{\rm VIII}:
\begin{eqnarray}
&{\rm VI})& \quad O_1^{\rm VI}=0.04 \pm 0.16. \nn \\
&{\rm VII})& \quad O_1^{\rm VII}=0.41 \pm 0.33. \nn \\
&{\rm VIII})& \quad O_2^{\rm VIII}=-0.48 \pm 0.37.  
\end{eqnarray}   

If we assume that future experimental results change and  indicate
that isospin breaking is indeed small, then, we can get an estimate of
${\cal
A}_{\rm CP}^{00} R_0$ ($O_1^{\rm VIII}$) from sum rule {\rm VIII} since
\begin{equation}
 \label{est1}
O_1^{\rm
VIII} \sim O_2^{\rm
VIII}. \end{equation}
We can do a similar exercise with sum rules {\rm VI} and {\rm 
VII}. Also using sum rule {\rm II} we can get an estimate
\begin{equation} \label{est2}
{\cal
A}_{\rm CP}^{00} R_0 \sim O_1^{\rm VI} - O_1^{\rm VII} + {\cal A}_{\rm 
CP}^{+0},
\end{equation}
that obviously would be in agreement with the estimate of 
eq.({\ref{est1}).         

\begin{figure}[t]
\vspace*{-1cm}
$$\hspace*{-1.cm}
\epsfysize=0.3\textheight
\epsfxsize=0.3\textheight
\epsffile{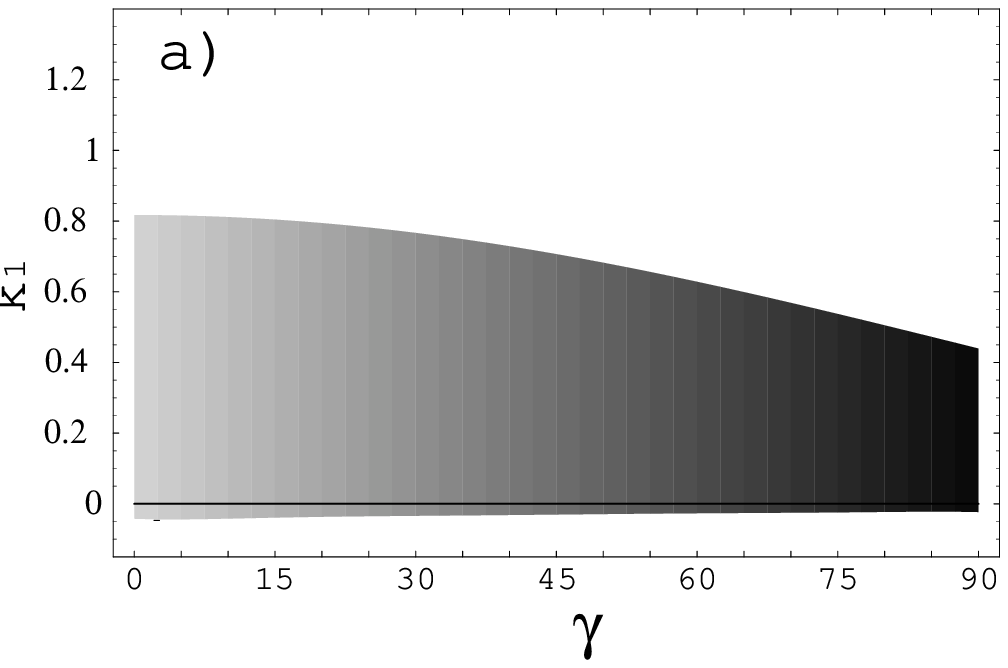} \hspace*{0.3cm}
\epsfysize=0.3\textheight
\epsfxsize=0.3\textheight
 \epsffile{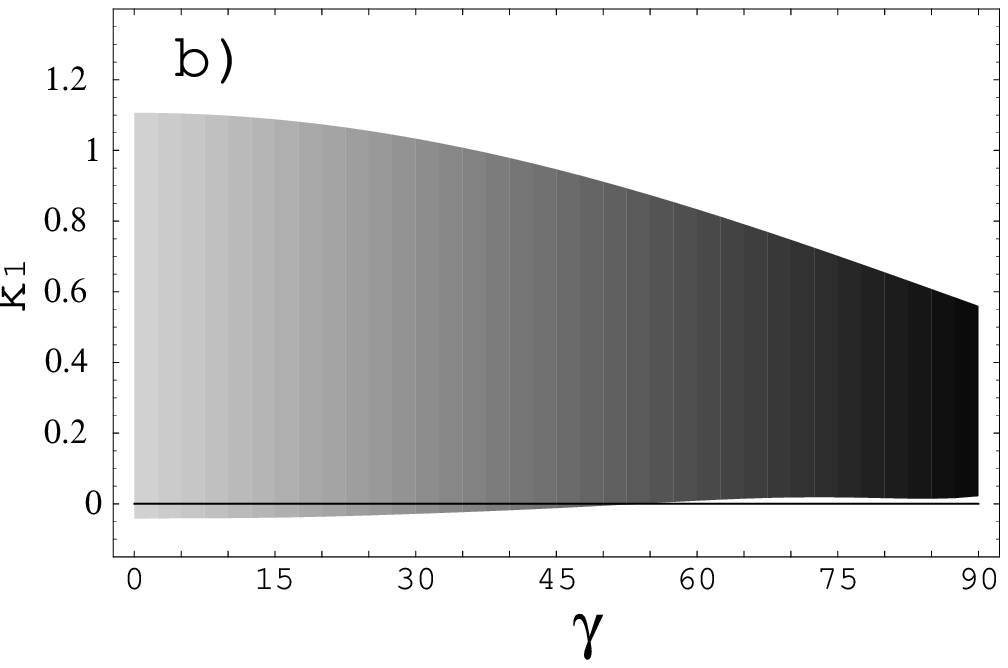}
$$
\vspace*{-0.9cm}
\caption[]{Sum rule {\rm I}: (a)  evaluated for the SM  but keeping 
free
the strong phases, (b) example of a model with $\xi_i=\xi_i^{\rm 
SM}$, free strong phases but $a_i^{\rm SM}/2 <a_i <2 a_i^{\rm SM}$ with 
no new 
weak 
phases ($b_i=0$). 
}\label{fig:fig9}                              
\end{figure}           

To end with the tests of the sum rules let's see what happens if 
one would like to use specific models  of New Physics. In this case it 
is 
necessary to evaluate for each 
model the new 
contributions to $\xi_{1,2}$, $a_{1,2}$, $b_{0,1,2}$ and to use
 the more general expressions for $k_1$ 
given in eq.(\ref{k1gen}),
$k_2$ in  eq.(\ref{k2gen}) and $u_\pm$ and $z_\pm$ as given in  
Appendix 
B. The expected main effect of
New Physics will be new contributions to the 
electroweak penguin parameters $a_1$, $a_2$ \cite{n1} ($q$ and 
$q_C$, respectively, in the Standard Model, see Appendix A) and 
possible new weak phases contributing to $b_0$, $b_1$ and $b_2$. 

We show in
Fig.\ref{fig:fig9}a  an example of the difference between the 
prediction 
for the SM of sum 
rule {\rm I}, using the general expression for $k_1$ 
eq.(\ref{k1gen}),
but this time keeping strong phases free (opposite to 
Fig.\ref{fig:fig1} 
where the 
strong phases are predicted from NLO QCD factorization) 
versus the prediction in Fig.\ref{fig:fig9}b of a generic 
model of New Physics that 
induces
important 
contributions to the electroweak penguins. This is useful to give us an 
idea of how important are the hadronic uncertainties coming from the 
strong phases. We see comparing Fig.\ref{fig:fig9}a with 
Fig.\ref{fig:fig1}a or Fig.\ref{fig:fig1}b 
that having 
a prediction for the strong phases changes dramatically the situation, 
however still there is an important region of Fig.\ref{fig:fig9}b non 
overlapping 
with Fig.\ref{fig:fig9}a. This is an 
example of a region that could only be explained by New Physics. This 
is useful to establish the line between new isospin violating 
physics and possible hadronic uncertainties coming from our model 
dependence in predicting strong 
phases.
In both cases, SM with free strong phases and New Physics, we  find a 
much 
better 
agreement with present 
experimental 
data for $k_1$ eq.({\ref{k1exp}}) than in Fig.\ref{fig:fig1}a-b. 
The reason of the decreasing behavior of $k_1$ with $\gamma$ can be 
understood analytically
taking  eq.(\ref{k1gen}) in the SM limit (Appendix A) 
 and observing that the maximal value of $k_1$ for 
$\gamma=\pi/2$ is approximately $(1+q)$ times smaller than the maximal 
$k_1$ for $\gamma=0$. (Notice that since in this approach the strong 
phase $\omega$  can 
take any value, 
the maximal value of $k_1$ corresponds precisely to $\omega \sim 
\pi$ 
opposite to the NLO QCD factorization prediction of $\omega \sim 0$).

In  supersymmetry, for instance, there are certain contributions that 
could be sizeable, in particular, those involving gluino-quark-squark
where a contribution of order $\alpha_s/m_{susy}^2$ can compete with 
Standard 
Model contributions of order $\alpha/M_W^2$ and can be as large as 
twice the SM predictions \cite{neu3}. 
Moreover, if New Physics contains new weak phases, they could 
contribute 
to $b_0$, $b_1$ and $b_2$.      
A global analysis of the  contributions from 
different models 
can be found in \cite{neu3}.

In conclusion, we have presented a set of sum rules 
that allow for an easy test of experimental data concerning the 
size of isospin 
breaking. They are derived in a model independent way but applied 
explicitly to 
the case of the Standard Model in the framework  
of NLO QCD factorization. Out of these sum rules a set 
of observables are proposed that permits a simple interpretation of 
data 
in terms of isospin breaking as a function of the  position
of the experimental point in their combined graphs. 
 The predicted results of these sum rules for the SM in QCD 
factorization are compared with
present data, showing in most cases  unexpectedly large central values
although with still too large experimental errors. It is of crucial 
interest to reduce these experimental errors to confirm or falsify the 
strong 
deviations from the SM predictions and to discern if experimental data 
fall 
in a non-overlapping region as in Fig.\ref{fig:fig9}a-b.

\bigskip
\centerline{\bf Acknowledgments}
\bigskip
It is a real pleasure to thank the theory group of Aachen: M. 
Beneke, W. Bernreuther, L. Sehgal, T. Feldmann,  K. 
Stergios, T. Teubner, N. D\"uchting 
and 
Sonya,  
for the warm 
hospitality and nice atmosphere during my stay there. Special thanks to 
Martin Beneke for very useful discussions and comments 
and also for understanding my personal decision, also 
to Werner 
Bernreuther for his generous support and help during my stay.   
I acknowledge financial support by BMBF during my 
stay in Germany and by Ministerio de Ciencia y Tecnologia in Spain.
\bigskip
\newpage


\newpage

\bigskip
\centerline{\bf Appendix A: Standard Model limit}
\bigskip   
\noindent
In order to make contact with the SM, we write down the 
parameters (amplitudes and phases) for the specific case of
the SM, using the notation of \cite{bbns99bis}\footnote{We understand 
$P$ in \cite{bbns99bis} as $P=|P| e^{i \phi_P}$. If one would 
like to make contact with the notation of \cite{n1} it is necessary to
identify our $\phi_T$ with $\tilde \phi$ of \cite{n1} and our
$\phi_T+\phi_P$ with $\phi_T$ of \cite{n1}.}

\begin{eqnarray} \label{smlimit}
&&\xi_{1} e^{i \theta_{1}}
\rightarrow {\epsilon_{3/2} e^{i(\phi_P + \phi + \pi)} \over 
\sqrt{1+\epsilon_{a}^2 {\rm cos}^2 \gamma
- 2 \epsilon_a
{\rm
cos} \gamma {\rm cos} \eta}}, \quad
a_{1} e^{i \phi_{a_{1}}} \rightarrow q e^{i \omega}, \quad
b_{1} e^{i \phi_{b_{1}}} \rightarrow 0,    \nn
\\
&&\xi_{2} e^{i \theta_{2}} \rightarrow {\epsilon_{\rm T} e^{i(\phi_P + 
\phi_{\rm T}+\pi)} \over
\sqrt{1+\epsilon_{a}^2 {\rm cos}^2 \gamma - 2 \epsilon_a
{\rm
cos} \gamma {\rm cos} \eta}}, \quad
a_{2} e^{i \phi_{a_{2}}} \rightarrow q_{C} e^{i \omega_{C}}, \quad
b_{2} e^{i \phi_{b_{2}}} \rightarrow 0,   \nn
\\
&&b_{0}  e^{i \phi_{b_{0}}} \rightarrow
{\epsilon_{\rm a} {\rm sin} \gamma 
e^{i(\eta -\delta)}
\over \sqrt{1+\epsilon_{a}^2 {\rm 
cos}^2 \gamma - 2 \epsilon_a
{\rm
cos} \gamma {\rm cos} \eta}}, \quad \theta_0 \rightarrow \phi_P+\delta,
\end{eqnarray}              
where
\begin{equation}
{\rm cos} \delta = {1 -\epsilon_a {\rm cos} \gamma {\rm cos} \eta \over
\sqrt{1+\epsilon_{a}^2 {\rm cos}^2 \gamma
- 2 \epsilon_a
{\rm
cos} \gamma {\rm cos} \eta}}, \quad 
{\rm sin} \delta ={ - \epsilon_a {\rm cos} \gamma {\rm sin} \eta \over
\sqrt{1+\epsilon_{a}^2 {\rm cos}^2 \gamma
- 2 \epsilon_a
{\rm
cos} \gamma {\rm cos} \eta}}.
\end{equation}

\bigskip
\centerline{\bf Appendix B: $u_{\pm}$, $z_{\pm}$ and $k_{1,2}$}
\bigskip     
\noindent
The dependence of $u_{\pm}=u \pm {\overline u}$ and 
$z_{\pm}=z \pm {\overline z}$ on $d_{1,2}$ is given by
\begin{eqnarray} \label{defomega}
{u}&=& 
 {2 \over x} \; {\rm Re} \left[ 
A \left(B^+\rightarrow 
\pi^+K^0 
\right) d_{2}^{*} \right]+
 {1 \over x} |{ d_{2}}|^{2},    
 \nn \\
{z}&=& 
 {2 \over x} \; {\rm Re} \left[ A 
\left(B^+\rightarrow
\pi^+K^0
\right) d_{1}^{*} \right]  +  {1 \over x} |{ d_{1}}|^{2}, 
\end{eqnarray}      
and the corresponding CP conjugates
\begin{eqnarray} \label{defomega2}
{\overline u}&=& 
 {2 \over x} \; {\rm Re} \left[ A 
\left(B^-\rightarrow
\pi^-{\overline K^0}
\right) {\overline d_{2}^{*}} \right] +{1 \over x} |{\overline 
d_{2}}|^{2},    \nn \\
{\overline z}&=& 
 {2 \over x} \; 
{\rm Re} \left[ A 
\left(B^-\rightarrow
\pi^-{\overline K^0}
\right) {\overline d_{1}^{*}} \right] + {1 \over x} |{\overline 
d_{1}}|^{2}.     
\end{eqnarray}     
The parameters $k_1$ and $k_2$ are
\begin{eqnarray} \label{defk1}
k_1 &=& {2 \over x} \; \left(
|{ d_{1}}|^2 + |{\overline { d_{1}}}|^2 - {\rm Re} [{ 
d_{1}} \; {d_{2}^*}] -
{\rm Re} [{\overline { d_{1}}} \; {\overline { 
d_{2}}^{*}}] \right),
 \\ \label{defk2}
k_2 &=& {2 \over x}\; \left(
|{ d_{1}}|^2 - |{\overline { d_{1}}}|^2 - {\rm Re} [{ 
d_{1}} \; 
{d_{2}^*}]
+
{\rm Re} [{\overline { d_{1}}} \; {\overline { d_{2}}^{*}}] 
\right), 
\end{eqnarray}    
with $x=2 (1+b_0^2) |P|^2$.
Notice that  the
dependence on $|P|$ cancels in
$u$, $z$, $k_1$ and $k_2$.  
From eqs.(\ref{defomega})-(\ref{defomega2}) and eq.(\ref{di}) we can 
derive a 
set of  model independent 
expressions for $u$, 
$\overline{u}$, $z$ and $\overline{z}$:
\begin{eqnarray} \label{omeganew}
u &=&
 {\xi_2 \over 1 +b_{0}^2} \;\left\{\;
 \; {\rm Re} \left[ e^{i (\theta_0 -\theta_2)
}
 \left(1-i b_0  e^{i
\phi_{b_{0}} } \right)  \left( e^{-i \gamma} - a_2 e^{-i
\phi_{a_{2}} }
+i b_2 e^{-i \phi_{b_{2}}} \right) \right] \right. 
\\&+& \left.
{\xi_{2} \over 2} \; \left[
(1+a_{2}^2+b_{2}^2)- 2 a_{2} \;{\rm cos}(\phi_{a_{2}}-\gamma) + 2 b_2 \;{\rm sin}
(\phi_{a_2}+ \gamma) -
2 a_2 b_{2} \; {\rm sin}(\phi_{a_{2}}-\phi_{b_2}) \right] \right\}. \nn
\end{eqnarray}          
Its CP conjugated $\overline{u}$ can be  obtained from 
eq.(\ref{omeganew})
by changing the sign of the weak phases: 
\begin{equation} \label{phaseschange}
\gamma \rightarrow - \gamma, \quad  \quad b_{0,2} \rightarrow 
-b_{0,2}.\end{equation}
$z$ is also obtained from eq.(\ref{omeganew})  substituting amplitudes 
and phases of $d_2$ by 
those of $d_1$: 
$$\xi_2 \rightarrow \xi_1, \quad
a_2 \rightarrow a_1,\quad b_2 \rightarrow b_1,\quad 
 \theta_2 \rightarrow \theta_1,\quad \phi_{a_2} \rightarrow \phi_{a_1},
 \quad   \phi_{b_2} 
\rightarrow 
\phi_{b_1}.$$
A similar  substitution to eq.(\ref{phaseschange}) will allow us to 
obtain 
the 
CP conjugate $\overline{z}$ from $z$
$$\gamma \rightarrow - \gamma, \quad \quad b_{0,1} \rightarrow 
-b_{0,1}.$$


\begin{thebibliography}{99}
\bigskip 

\bibitem{fl} R.~Fleischer, {\it Phys. Lett.} {\bf B365} (1996) 399;
{\it Phys. Lett.} {\bf B459} (1999) 306
M. Neubert and J.R. Rosner, {\it Phys. Rev. Lett.} {\bf 81} (1998) 
5076; {\it Phys. Lett.} {\bf B441} (1998) 403.

\bibitem{fl2} A.J.~Buras and 
R.~Fleischer, {\it Eur. Phys. J.} {\bf C11} (1999) 93; 
{\it Eur. Phys. J.} {\bf C16} (2000) 97.

\bibitem{n1} M. Neubert, {\it JHEP} {\bf 02} (1999) 014.


\bibitem{bbns99}
M.~Beneke, G.~Buchalla, M.~Neubert, C.T.~Sachrajda,
{\it Phys. Rev. Lett.} {\bf 83}, (1999) 1914; {\it Nucl. Phys.} {\bf 
B591} (2000) 313. See also: M.~Neubert, hep-ph/0012204.

\bibitem{bbns99bis} M.~Beneke, G.~Buchalla, M.~Neubert, C.T.~Sachrajda,    
 hep-ph/007256.

\bibitem{bbns00}
M.~Beneke, G.~Buchalla, M.~Neubert, C.T.~Sachrajda, hep-ph/0104110.
   

\bibitem{mu}
T.~Muta, A.~Sugamoto, M.Z.~Yang and Y.D.~Yang, {\it Phys. Rev.} {\bf 
D62} (2000) 094020;
M.Z.~Yang and Y.D.~Yang, {\it Phys. Rev.} {\bf D62} (2000) 114019;
M.Z.~Yang and Y.D.~Yang, hep-ph/0012208;
X.G.~He, J.P.~Ma and C.Y.~Wu, {\it Phys. Rev.} {\bf D63} (2001) 094004.

\bibitem{silv} M.~Ciuchini, E.~Franco, G.~Martinelli, M.~Pierini, 
L.~Silvestrini, hep-ph/0104018.

\bibitem{keum} Y.Y.~Keum, H.-n~Li and A.I.~Sanda, hep-ph/0004004; {\it 
Phys. Rev.} {\bf D63} (2001) 054008; Y.Y. Keum and H.-n.~Li, {\it Phys.
Rev.} {\bf D63} (2001) 074006.

\bibitem{gr3} R.~Fleischer, {\it Phys. Lett.} {\bf B435} (1998) 221;
M.~Neubert, JHEP {\bf 9902} (1999) 014; hep-ph/9910530; 
hep-ph/9904321; {\it Nucl.Phys.Proc.Suppl.} {\bf 
86} (2000) 477;      
M.~Gronau, {\it Phys. Rev.} {\bf D62} (2000) 014031;
X.-G. He, C.-L. Hsueh, J.-Q. Shi, {\it Phys. Rev. Lett.} {\bf 84} 
(2000) 18;  Zhi-zhong Xing, {\it Phys.Lett.} {\bf B488} (2000) 162; 
X.G. He, Y.K. Hsiao, J.Q. Shi, Y.L. Wu, 
Y.F. Zhou, {\it Phys.Rev.} {\bf D62} (2000) 036007;
hep-ph/0011337; A. F. Falk, A. A. Petrov, 
{\it Phys.Rev.Lett.} {\bf 85} (2000) 252;
R.E. Blanco, C. Gobel, R. Mendez-Galain,  hep-ph/0007105; 
C.S. Kim, S. Oh, hep-ph/0009082; hep-ph/0103319.






\bibitem{jo} R.~Fleischer and J.~Matias, {\it Phys. Rev.} 
{\bf D61} (2000) 074004. 

\bibitem{neu3} Y.~Grossman, M.~Neubert, A.L.~Kagan, JHEP 9910 
(1999) 029. 

\bibitem{exp} D.Cronin-Hennessy et al. [CLEO Collaboration], 
hep-ex/0001010; D.M. Asner et al. [CLEO Collaboration], hep-ex/0103040;
G. Cavoto, [Babar Collaboration], talk given at the XXXVI Rencontres de 
Moriond QCD; B. Casey [Belle Collaboration],  talk given at the XXXVI 
Rencontres de Moriond QCD.

\bibitem{exp2} S.~Chen et al. [CLEO Collaboration], {\it Phys. Rev. 
Lett.} {\bf 85} (2000) 525. 

\bibitem{ball} See the report B Decays at LHC, P. Ball et al., 
hep-ph/0003238.

\bibitem{lip} H.J.~Lipkin, hep-ph/9809347.

\bibitem{gron} M.~Gronau and J.L.~Rosner, {\it Phys. Rev.}
{\bf D59} (1999) 113002.

\bibitem{bbl} G.~Buchalla, A.J.~Buras and M.E.~Lautenbacher, {\it Rev. 
Mod. Phys.} {\bf 68} (1996) 1125.

\bibitem{ciu} M.~Ciuchini et. al, hep-ph/0012308.

\end{thebibliography}
\end{document}